\documentclass[aps,prd,twocolumn,superscriptaddress,showpacs,
letterpaper]{revtex4}
\usepackage{graphicx}
\usepackage{dcolumn}
\usepackage{bm}

\def\met{\mathbin{E\mkern - 11mu/_T}}

\def\GeV{\mathbin{\hbox{GeV}}}
\def\GeVc{\mathbin{\hbox{GeV}\mkern - 3mu/\mkern -1mu c}}
\def\GeV2c{\mathbin{\hbox{GeV}\mkern - 3mu/\mkern -1mu c^2}}
\def\TeV2c{\mathbin{\hbox{TeV}\mkern - 3mu/\mkern -1mu c^2}}
\def\MeV2c{\mathbin{\hbox{MeV}\mkern - 3mu/\mkern -1mu c^2}}
\def\lesim{\mathrel {\vcenter {\baselineskip 0pt \kern 0pt
\hbox{$<$} \kern 0pt \hbox{$\sim$} }}}
\def\gesim{\mathrel {\vcenter {\baselineskip 0pt \kern 0pt
\hbox{$>$} \kern 0pt \hbox{$\sim$} }}}

\begin{document}

\title{
Search for {\boldmath $W^\prime$} Boson Decaying to Electron-Neutrino
Pairs in 
{\boldmath $p\bar p$} Collisions at {\boldmath $\sqrt{s}=1.96$}~TeV 
\\
}

\affiliation{Institute of Physics, Academia Sinica, Taipei, Taiwan 11529, 
Republic of China} 
\affiliation{Argonne National Laboratory, Argonne, Illinois 60439} 
\affiliation{Institut de Fisica d'Altes Energies, Universitat Autonoma de 
Barcelona, E-08193, Bellaterra (Barcelona), Spain} 
\affiliation{Baylor University, Waco, Texas  76798} 
\affiliation{Istituto Nazionale di Fisica Nucleare, University of Bologna, 
I-40127 Bologna, Italy} 
\affiliation{Brandeis University, Waltham, Massachusetts 02254} 
\affiliation{University of California, Davis, Davis, California  95616} 
\affiliation{University of California, Los Angeles, Los Angeles, California  
90024} 
\affiliation{University of California, San Diego, La Jolla, California  92093} 
\affiliation{University of California, Santa Barbara, Santa Barbara, California
93106} 
\affiliation{Instituto de Fisica de Cantabria, CSIC-University of Cantabria, 
39005 Santander, Spain} 
\affiliation{Carnegie Mellon University, Pittsburgh, PA  15213} 
\affiliation{Enrico Fermi Institute, University of Chicago, Chicago, Illinois 
60637} 
\affiliation{Joint Institute for Nuclear Research, RU-141980 Dubna, Russia} 
\affiliation{Duke University, Durham, North Carolina  27708} 
\affiliation{Fermi National Accelerator Laboratory, Batavia, Illinois 60510} 
\affiliation{University of Florida, Gainesville, Florida  32611} 
\affiliation{Laboratori Nazionali di Frascati, Istituto Nazionale di Fisica 
Nucleare, I-00044 Frascati, Italy} 
\affiliation{University of Geneva, CH-1211 Geneva 4, Switzerland} 
\affiliation{Glasgow University, Glasgow G12 8QQ, United Kingdom} 
\affiliation{Harvard University, Cambridge, Massachusetts 02138} 
\affiliation{Division of High Energy Physics, Department of Physics, University
of Helsinki and Helsinki Institute of Physics, FIN-00014, Helsinki, Finland} 
\affiliation{University of Illinois, Urbana, Illinois 61801} 
\affiliation{The Johns Hopkins University, Baltimore, Maryland 21218} 
\affiliation{Institut f\"{u}r Experimentelle Kernphysik, Universit\"{a}t 
Karlsruhe, 76128 Karlsruhe, Germany} 
\affiliation{High Energy Accelerator Research Organization (KEK), Tsukuba, 
Ibaraki 305, Japan} 
\affiliation{Center for High Energy Physics: Kyungpook National University, 
Taegu 702-701, Korea; Seoul National University, Seoul 151-742, Korea; and 
SungKyunKwan University, Suwon 440-746, Korea} 
\affiliation{Ernest Orlando Lawrence Berkeley National Laboratory, Berkeley, 
California 94720} 
\affiliation{University of Liverpool, Liverpool L69 7ZE, United Kingdom} 
\affiliation{University College London, London WC1E 6BT, United Kingdom} 
\affiliation{Centro de Investigaciones Energeticas Medioambientales y 
Tecnologicas, E-28040 Madrid, Spain} 
\affiliation{Massachusetts Institute of Technology, Cambridge, Massachusetts  
02139} 
\affiliation{Institute of Particle Physics: McGill University, Montr\'{e}al, 
Canada H3A~2T8; and University of Toronto, Toronto, Canada M5S~1A7} 
\affiliation{University of Michigan, Ann Arbor, Michigan 48109} 
\affiliation{Michigan State University, East Lansing, Michigan  48824} 
\affiliation{Institution for Theoretical and Experimental Physics, ITEP, 
Moscow 117259, Russia} 
\affiliation{University of New Mexico, Albuquerque, New Mexico 87131} 
\affiliation{Northwestern University, Evanston, Illinois  60208} 
\affiliation{The Ohio State University, Columbus, Ohio  43210} 
\affiliation{Okayama University, Okayama 700-8530, Japan} 
\affiliation{Osaka City University, Osaka 588, Japan} 
\affiliation{University of Oxford, Oxford OX1 3RH, United Kingdom} 
\affiliation{University of Padova, Istituto Nazionale di Fisica Nucleare, 
Sezione di Padova-Trento, I-35131 Padova, Italy} 
\affiliation{LPNHE, Universite Pierre et Marie Curie/IN2P3-CNRS, UMR7585, 
Paris, F-75252 France} 
\affiliation{University of Pennsylvania, Philadelphia, Pennsylvania 19104} 
\affiliation{Istituto Nazionale di Fisica Nucleare Pisa, Universities of Pisa, 
Siena and Scuola Normale Superiore, I-56127 Pisa, Italy} 
\affiliation{University of Pittsburgh, Pittsburgh, Pennsylvania 15260} 
\affiliation{Purdue University, West Lafayette, Indiana 47907} 
\affiliation{University of Rochester, Rochester, New York 14627} 
\affiliation{The Rockefeller University, New York, New York 10021} 
\affiliation{Istituto Nazionale di Fisica Nucleare, Sezione di Roma 1, 
University of Rome ``La Sapienza," I-00185 Roma, Italy} 
\affiliation{Rutgers University, Piscataway, New Jersey 08855} 
\affiliation{Texas A\&M University, College Station, Texas 77843} 
\affiliation{Istituto Nazionale di Fisica Nucleare, University of Trieste/\ 
Udine, Italy} 
\affiliation{University of Tsukuba, Tsukuba, Ibaraki 305, Japan} 
\affiliation{Tufts University, Medford, Massachusetts 02155} 
\affiliation{Waseda University, Tokyo 169, Japan} 
\affiliation{Wayne State University, Detroit, Michigan  48201} 
\affiliation{University of Wisconsin, Madison, Wisconsin 53706} 
\affiliation{Yale University, New Haven, Connecticut 06520} 
\author{A.~Abulencia}
\affiliation{University of Illinois, Urbana, Illinois 61801}
\author{J.~Adelman}
\affiliation{Enrico Fermi Institute, University of Chicago, Chicago, Illinois 
60637}
\author{T.~Affolder}
\affiliation{University of California, Santa Barbara, Santa Barbara, California
93106}
\author{T.~Akimoto}
\affiliation{University of Tsukuba, Tsukuba, Ibaraki 305, Japan}
\author{M.G.~Albrow}
\affiliation{Fermi National Accelerator Laboratory, Batavia, Illinois 60510}
\author{D.~Ambrose}
\affiliation{Fermi National Accelerator Laboratory, Batavia, Illinois 60510}
\author{S.~Amerio}
\affiliation{University of Padova, Istituto Nazionale di Fisica Nucleare, 
Sezione di Padova-Trento, I-35131 Padova, Italy}
\author{D.~Amidei}
\affiliation{University of Michigan, Ann Arbor, Michigan 48109}
\author{A.~Anastassov}
\affiliation{Rutgers University, Piscataway, New Jersey 08855}
\author{K.~Anikeev}
\affiliation{Fermi National Accelerator Laboratory, Batavia, Illinois 60510}
% ...
\collaboration{CDF Collaboration}
\noaffiliation

\begin{abstract}
We present the results of a search for $W^\prime$ boson decaying
to electron-neutrino pairs in $p\bar p$ collisions at a center-of-mass
energy of 1.96 TeV, using a data sample corresponding to $205~\rm{pb}^{-1}$
of integrated luminosity collected by the CDF II detector at Fermilab.
We observe no evidence for this decay mode and set limits on the production
cross section times branching fraction, assuming the neutrinos from
$W^\prime$ boson decays to be light. If we assume the manifest left-right
symmetric model, we exclude a $W^\prime$ boson with mass less than
$788~\GeV2c$ at the 95\% confidence level.
\\ \vspace*{0.5cm}
\end{abstract}
\maketitle
\date{\today}

Although to date all data are consistent with the standard model 
of particle physics, the model is not a complete theory. 
For example, it does not explain the number of lepton and quark 
generations nor their mass hierarchy. 
Many theories have been proposed to address these deficiencies
of the standard model. Some of these theories contain gauge 
symmetries that can be spontaneously broken down to the left-right 
symmetry~\cite{higher symm} featuring a right-handed SU(2) symmetry 
and corresponding additional gauge bosons, including a right-handed 
charged heavy vector boson, generically known as a $W^\prime$ 
boson~\cite{left_right,prime_model}.

Previous direct searches for a new charged heavy vector boson have set 
model-dependent limits on the cross section times branching fraction.
Searches considering the decay mode $W^\prime\to e\nu_e$ 
and $W^\prime\to \mu\nu_\mu$ have excluded a $W^\prime$ boson with
a mass below $754$ and $660~\GeV2c$, respectively, at the 95\% confidence 
level (CL)~\cite{wprime_enu,wprime_munu}. 
Assuming the universality of
lepton-$W^\prime$ boson couplings, a $W^\prime$ boson with a mass 
below $786~\GeV2c$ has been excluded at the 95\% CL
by combining the limits of both leptonic decay modes~\cite{wprime_enu}. 
Also, a search considering a decay mode  $W^\prime\to t\bar b$ has excluded
at the 95\% CL a $W^\prime$ boson with a mass
below $670~\GeV2c$ for models with a right-handed neutrino that is heavier 
than a $W^\prime$ boson~\cite{wprime_tb}.
These mass limits all assume the manifest left-right symmetry, where
the right-handed CKM matrix and the gauge coupling constant are
identical to those of the standard model~\cite{lrmodel}.
Indirect searches have set model independent mass limits with less 
sensitivity studying, for example, the Michel spectrum in polarized muon 
decay~\cite{wr_mass}.

In this report, we present the results of a search for a $W^\prime$ 
boson in the $e\nu_e$ decay mode~\cite{charge conjugate}. We use a 
data sample with an integrated 
luminosity of $205$~pb$^{-1}$ of $p\bar p$ collisions at 
$\sqrt{s}=1.96$~TeV recorded by the upgraded Collider Detector 
at Fermilab (CDF II) during 2002--2003. 
This search is based on an analysis of high mass $e\nu_e$ final 
state candidates and assumes the neutrino from a $W^\prime$ boson 
decay to be light and stable.

In this search, we select events that are consistent with the 
production of the standard model $W$ boson followed by its decay to 
an $e\nu_e$ final state and any heavier object that decays in the 
same manner. 
We set a limit on the production and decay of the heavier objects 
normalized by the observed rate of W bosons to this final state.
This technique allows us to avoid uncertainties associated with the 
measurement of the absolute cross section or limit. This limit on the 
rate can be applied to any process that yields an $e\nu_e$ final state 
and has similar kinematic properties. We then use the limits on the 
rates to set a limit on the mass of the $W^\prime$ boson,
assuming the manifest left-right symmetric model and a 
suppression of diboson decay channels.  
The CDF II detector is described in detail elsewhere~\cite{CDF_General}.
We use a coordinate system where $\theta$ is the polar angle to the proton 
beam, $\phi$ is the azimuthal angle about the beam axis, and $\eta$ is the
pseudorapidity defined as $-\ln(\tan(\theta/2))$.
The detector has a charged particle tracking system immersed in a 1.4-T 
solenoidal magnetic field coaxial with the proton and antiproton beams. 
The tracking system consists of an open-cell drift chamber surrounding 
a silicon tracking system that measures particle momentum. The 
electromagnetic and hadronic calorimeters surrounding the 
tracking system measure the energy of particles that interact 
electromagnetically or hadronically. These calorimeters are segmented 
in a projective tower geometry and divided into central calorimeters
covering $|\eta|<1.1$ and forward calorimeters covering 
$1.2<|\eta|<3.6$. 
An electron candidate is identified by an energy deposit with a track 
pointing to it in the electromagnetic calorimeter. 
A set of charged particle detectors 
surrounding the calorimeters is used to identify muon candidates with 
$|\eta|<1.0$.

Candidate events are identified by the CDF trigger system requiring 
at least one electron candidate in the central electromagnetic calorimeter 
with transverse energy $E_T> 18$~GeV and a matching track with transverse 
momentum $p_T>9~\GeVc$, where $E_T$ and $p_T$ are energy and momentum
measured transverse to the beam line, respectively.
An additional trigger with $E_T>70$~GeV and no restriction on 
the amount of energy leakage into the hadronic calorimeter
is used to ensure high efficiency for high $E_T$ electrons.
Subsequently, we refine the candidate sample after full event 
reconstruction by requiring an electron candidate with 
$E_T>25$~GeV and its track 
$p_T$ greater than $15~\GeVc$ in the fiducial region of the detector 
within $|\eta|<1.0$. 
We also require the electron candidates to be well isolated from 
energy flow in the event and to have shower profiles consistent 
with that of electron initiated showers~\cite{shower profile}.
The presence of a neutrino is inferred from a sizable missing
transverse energy, $\met$~\cite{met}. 
We require the missing transverse energy in the event, $\met$,
to be greater than 25~GeV.

Additional requirements are imposed to reject specific sources of 
background. Dilepton events from Drell-Yan, $t\bar{t}$, 
and diboson backgrounds are removed by vetoing events with a 
second isolated lepton candidate, either an electron or a muon, 
with $p_T > 15~\GeVc$.
QCD multijet events get misclassified into the $W/W^\prime\to e\nu_{e}$ 
sample when one of the jets is misidentified as an electron candidate and
the $\met$ requirement is satisfied due to energy mismeasurement.
In these QCD multijet events, the $\met$ due to jet energy mismeasurement
is generally much smaller than the $E_T$ of the jet misidentified as an 
electron candidate when the $E_T$ is large.
However, $W/W^\prime\to e\nu_e$ events will produce $E_T$ and $\met$ 
comparable in magnitude, if the $p_T$ of the boson is much smaller
than the mass of the boson.
We require the ratio of the electron candidate $E_T$ to $\met$ to 
be between 0.4 and 2.5.
Any $W/W^\prime\to e\nu_{e}$ events that lie outside the allowed 
region mostly have a high $p_T$ boson. This requirement has an 
efficiency above 99\% for $W/W^\prime\to e\nu_{e}$ events
and an estimated rejection rate of $\sim 40$\% for 
the misclassified QCD multijet events with high $E_T$.  
Additional details of the event selection requirements are
presented in Ref.~\cite{thesis}.

\begin{figure}[htbp]
\begin{center}
\includegraphics[width=3.5in]{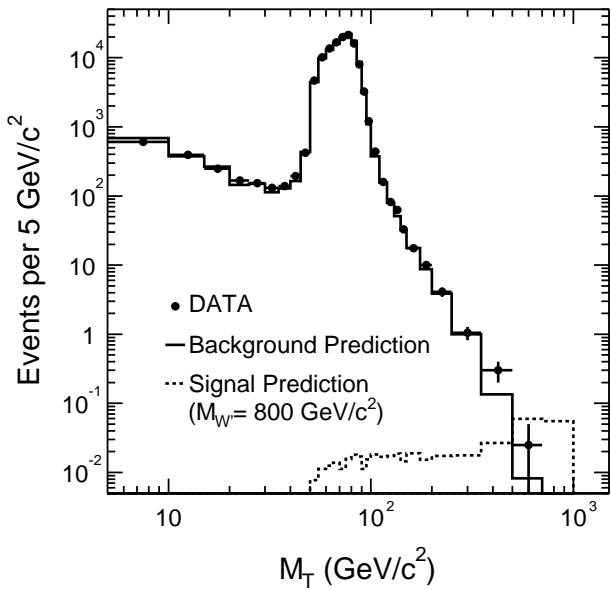}
\caption{The transverse mass distributions of $e\nu_e$ candidate events.
The background rates are obtained from the fit described in the text.
The distribution expected from the production of a $W^\prime$ boson of
$M_{W^\prime}=800~\GeV2c$ is shown by the dashed line.}
\label{fig:mtfit}
\end{center}
\end{figure}

The resulting sample contains 120\,484 events. The transverse mass of a 
candidate event is calculated as 
\begin{equation}
M_T\equiv\sqrt{2E_T\met(1-\cos\phi_{e\nu})},
\end{equation} 
where $\phi_{e \nu}$ is the azimuthal opening angle between the 
electron candidate and the $\met$ direction. This $M_T$ distribution has a 
clear Jacobian peak associated with the production and decay of 
the $W$ boson as shown in Fig.~\ref{fig:mtfit}. 

The shapes of $M_T$ distributions and the acceptance times efficiency 
for the $W^\prime$ boson signal are estimated using {\sc pythia} Monte 
Carlo calculation~\cite{pythia} with CTEQ5L parton distribution functions 
(PDFs)~\cite{ref:cteq}, together with the {\sc geant}~\cite{geant}--based 
CDF II detector simulation. 
We require the $W^\prime$ boson to have right-handed couplings to the 
fermions when generating signal Monte Carlo samples. The acceptance times 
efficiency rises from 41\% for  
$M_{W^\prime}=200~\GeV2c$, plateaus at 48\% for $350~\GeV2c$, remains roughly
flat up to $800~\GeV2c$, and then falls to 45\% for $M_{W^\prime}=950~\GeV2c$.
The initial increase in the acceptance times efficiency is due to a heavier 
$W^\prime$ boson produced more centrally. The subsequent fall is due to event 
selection requirements becoming less efficient for the high-energy electrons. 
We use a next-to-next-to-leading order (NNLO) cross section 
prediction~\cite{hamberg} for the $W^\prime$ boson production using MRST1 
PDFs~\cite{ref:MRST}.

The largest background sources are $W$ or $Z$ boson production with the 
boson decaying to final states that contain electrons. These include 
$W\to e\nu_e$ which is the dominant background, 
$W\to \tau\nu_{\tau}\to e\nu_e\nu_{\tau}\nu_{\tau}$, 
$Z/\gamma^*\to ee$, and $Z/\gamma^*\to\tau\tau\to eX$. 
The other background sources are electrons coming from diboson production 
and $t\bar t$ production, and jets misidentified as electron candidates 
from QCD multijet production.
The shapes of the $M_T$ distributions and acceptance times efficiency of
the non-multijet backgrounds are calculated using samples generated 
with  {\sc pythia}, except for $WW$ and $WZ$ backgrounds, which are
calculated with {\sc alpgen}~\cite{alpgen}, interfaced with 
{\sc herwig}~\cite{herwig}. All Monte Carlo samples are subjected to the 
CDF II detector simulation.
We use theoretical cross section predictions to estimate 
the expected background yields~\cite{hamberg,ref:nlo,ref:tt}. 

For the QCD multijet background, the predicted number of events is 
estimated from the data sample using the azimuthal opening angle
between the direction of the electron 
candidate and the vector sum of the jet energy flow. Since a jet 
misidentified as an electron candidate will be seen as recoiling 
against the rest of the jets in the event, we expect to see back-to-back 
behavior in the azimuthal opening angle, whereas $W/W^\prime\to e\nu_e$ 
events do not have a strong angular correlation~\cite{thesis}.
The data and estimated background $M_T$ distributions
are compared in Fig.~\ref{fig:mtfit}. A small 
excess of events with a significance of about 1.8 standard deviations 
above the background 
expectation is observed in the 350--500~$\GeV2c$ $M_T$ bin. 
The contributions from $W\to e\nu_e$, QCD multijet, and the rest of the 
backgrounds above $M_T=200~\GeV2c$ are listed in Table~\ref{tab:nfit_bin}.

\begin{table*}[htbp]
\caption{The event yields for the background sources in $M_T$ above
$200~\GeV2c$ compared to the observed data. The uncertainties are 
correlated. The correlations are properly taken care of in the 
systematic uncertainty estimation.}
\begin{center}
\def\arraystretch{1.0}
\begin{tabular}{cccccc}\hline\hline 
Background        &\multicolumn{5}{c}{Events in each $M_T$ bin ($\GeV2c$)} \\
    &200 - 250     &250 - 350     &350 - 500      &500 - 700      
&700 - 1000     \\\hline
$W \to e\nu$      &$30.8\pm 5.7$ &$17.0\pm 4.0$ &$3.52\pm 1.70$ 
&$0.27\pm 0.45$ &$<0.01$ \\
QCD Multijet          &$ 2.7\pm 6.1$ &$ 0.0\pm 3.3$ 
&$0.00\pm 0.29$ &$0.00\pm 0.01$ &$<0.01$ \\
Other Backgrounds &$ 5.2\pm 1.0$ &$ 3.0\pm 0.9$ &$0.51\pm 0.22$ 
&$0.06\pm 0.08$ &$0.00\pm 0.03$ \\\hline
Total Background  &$38.7\pm 8.9$ &$20.0\pm 5.9$ &$4.03\pm 1.97$ 
&$0.33\pm 0.53$ &$0.00\pm 0.03$ \\\hline
Data              &$41$          &$21$          &$9$            
&$1$            &$0$            \\\hline\hline
\end{tabular}
\end{center}
\label{tab:nfit_bin}
\end{table*}

In order to estimate the size of the potential signal contribution 
in the sample, a binned maximum likelihood fit is performed on the 
observed $M_T$ distribution between 0 and 1\,500~$\GeV2c$, using the 
background predictions and the expected $W^\prime$ boson contribution 
with different mass values ranging from 200 to 950~$\GeV2c$. 
The fit results are shown in Table~\ref{tab:valuefit}, expressed as 
\begin{equation}
\beta \equiv {\sigma\cdot{\cal B}(W^\prime\to e\nu_e)\over
\sigma\cdot{\cal B}(W^\prime\to e\nu_e)_{LR}}, 
\end{equation}
where the numerator is the observed cross section times branching 
fraction and the denominator is the expected value from the manifest 
left-right symmetric model.
The expected signal yield is normalized by the observed $W$ boson yield
obtained from the fit. 
This normalization reduces the effects of uncertainties common to both 
the $W$ boson and W' boson yields, such as the uncertainties of an 
integrated luminosity and of theoretical cross sections.

\begin{table}[htbp]
\begin{center}
\end{center}
\caption{The expected numbers of the events from the 
$W^\prime\to e\nu_e$ process, $N_{\rm{exp}}$, assuming the manifest 
left-right symmetric model and normalized by the observed $W$ boson 
yield.  We also show observed rate of $W^\prime$ boson production 
from the fit described in the text, and the 95\% CL upper limit on 
this rate. The uncertainties are statistical only and do not include 
systematic uncertainties. The 95\% upper limits include both statistical 
and systematic uncertainties.}
\def\arraystretch{1.0}
\begin{tabular}{cccc}
\hline\hline
$M_{W^\prime}$   &$N_{\rm{exp}}$ 
&\multicolumn{2}{c}{$\beta \left(={\sigma\cdot{\cal B}
(W^\prime\to e\nu_e)\over 
\sigma\cdot{\cal B}(W^\prime\to e\nu_e)_{LR}}\right)$} \\
$(\GeV2c)$  &(events)  &Fit		&Upper Limit   \\\hline 
200     &12000         &${0.002_{-0.001}^{+0.002}}$ &0.01 \\
250     &5390          &${0.002_{-0.002}^{+0.002}}$ &0.01 \\
300     &2630          &${0.001_{-0.001}^{+0.003}}$ &0.01 \\
350     &1380          &${0.008_{-0.005}^{+0.006}}$ &0.02 \\
400     &753           &${0.012_{-0.006}^{+0.008}}$ &0.03 \\
450     &417           &${0.019_{-0.010}^{+0.012}}$ &0.03 \\
500     &236           &${0.028_{-0.015}^{+0.019}}$ &0.08 \\
550     &135           &${0.030_{-0.021}^{+0.030}}$ &0.11 \\
600     &77.0          &${0.034_{-0.026}^{+0.026}}$ &0.16 \\
650     &43.7          &${0.040_{-0.036}^{+0.064}}$ &0.23 \\
700     &25.1          &${0.050_{-0.050}^{+0.099}}$ &0.36 \\
750     &14.2          &${0.060_{-0.060}^{+0.162}}$ &0.57 \\
800     &7.89          &${0.075_{-0.075}^{+0.279}}$ &1.00 \\
850     &4.35          &${0.081_{-0.081}^{+0.512}}$ &1.84 \\
900     &2.35          &${0.06_{-0.06}^{+1.00}}$ &3.70  \\
950     &1.24          &${0.00_{-0.00}^{+2.15}}$ &8.26  \\
\hline\hline
\end{tabular}
\label{tab:valuefit}
\end{table}

We set upper limits on the rate of a $W^\prime$ boson
by constructing the posterior distribution for $\beta$ for each fixed 
value of $M_{W^\prime}$. The likelihood is maximized for a fixed value 
of $\beta$ with respect to the background contributions. 
We use the resulting likelihood distribution to set the 95\% CL upper 
limit on the ratio $\beta$ by numerically integrating over $\beta$.
We consider the likelihood function only in the physical region
where $\beta$ is greater than or equal to zero. 
Systematic uncertainties in the signal and background rates are 
incorporated in the upper limit using the Bayesian prescription of 
convoluting the likelihood function with a truncated Gaussian prior 
distribution for each nuisance parameter~\cite{Luc}.
The upper limits in the cross section times branching fraction are 
obtained by multiplying the upper limits in $\beta$ by the theoretical 
cross section times branching fraction.

\begin{figure}[htbp]
\begin{center}
\includegraphics[width=3.5in]{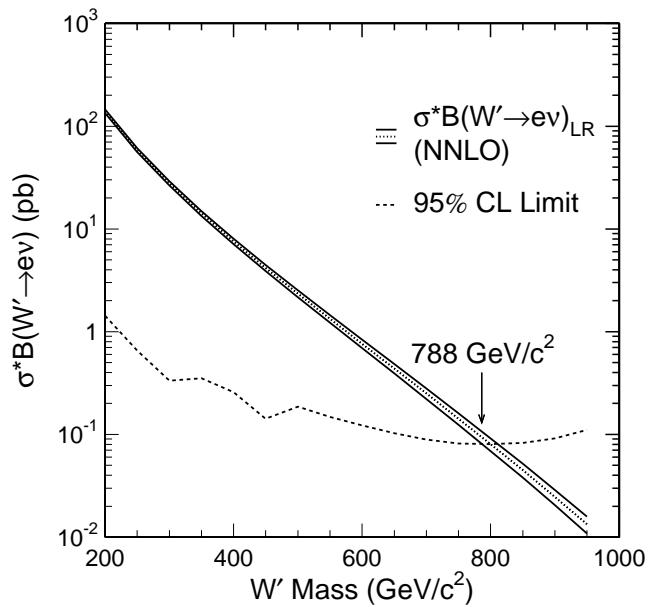}
\caption{The 95\% CL limits on cross section times branching fraction
as a function of $W^\prime$ boson mass. The region above the dashed line is 
excluded at the 95\% CL. Also, the cross section times branching fraction 
assuming the manifest left-right symmetric model, 
$\sigma\cdot{\cal B}(W^\prime\to e\nu_e)_{LR}$, is shown along with 
its uncertainty. The intercept of the cross section limit curve and
the lower bound of the theoretical cross section yields 
$M_{W^\prime}>788~\GeV2c$ at the 95\% CL.}
\label{fig:limits}
\end{center}
\end{figure}

We consider systematic uncertainties due to uncertainties in PDFs, 
electron energy measurement, initial state radiation, and jet energy 
measurement.  Since the uncertainty in the theoretical cross section 
of the $W^\prime$ boson production does not affect the limit on the 
cross section, this uncertainty is not included when calculating the 
limit. The systematic uncertainty in the PDFs is decomposed into a 
component affecting the theoretical $W^\prime$ production cross 
section and one affecting the expected acceptance of the $W^\prime$
boson. The component affecting the theoretical cross section is thus 
removed from the cross section limit calculation but contributes to 
the mass limit calculation. The cross section uncertainties of 
backgrounds are also taken into account.
The largest contribution to the systematic uncertainty below 
$M_{W^\prime} = 700~\GeV2c$ comes from the uncertainty in the 
jet energy measurement. 
The uncertainty in the signal acceptance due to the uncertainty in the 
PDFs is the dominant contribution above $700~\GeV2c$.
The resulting upper limits in $\beta$ are summarized in 
Table~\ref{tab:valuefit} and the upper limits in the production 
cross section times branching fraction
are plotted as a function of $M_{W^\prime}$ in Fig.~\ref{fig:limits}. 

Using theoretical predictions assuming the manifest left-right symmetric 
model, which has the right-handed CKM matrix and the gauge coupling 
constant identical to those of the standard model, these limits on the 
cross section times branching fraction are converted into limits on 
the mass of a $W^\prime$ boson. The uncertainty on the theoretical 
cross section is calculated by varying the momentum transfer scale and 
by using the uncertainties in the PDF parametrization. The uncertainty 
is shown as the band in Fig.~\ref{fig:limits}. We take the lower bound 
of the theoretical cross section to obtain the mass limit. This allows 
us to exclude a $W^\prime$ boson with mass below $788~\GeV2c$ at the 
95\% CL.

We estimate the sensitivity of our search to be $835~\GeV2c$ by 
calculating the median expected mass limit in a large ensemble of 
background-only pseudo-experiments. The discrepancy between the 
measured limit and the expected limit is largely due to the excess, 
though not statistically significant, observed in the $M_T$ region 
of 350--500~$\GeV2c$. 

In summary, we have performed a search for a new heavy charged vector 
boson decaying to an electron-neutrino pair with a light and stable 
neutrino in 1.96~TeV $p\bar p$ collisions. 
We do not observe any statistically significant excess over background 
expectations. We use a fit of the $M_T$ distribution to set upper limits 
on the production and decay rate of a $W^\prime$ boson
and exclude a $W^\prime$ boson with $M_{W^\prime}<788~\GeV2c$ at 
the 95\% CL, assuming the manifest left-right symmetric model. 

We thank the Fermilab staff and the technical staffs of the participating
institutions for their vital contributions. This work was supported by the
U.S. Department of Energy and National Science Foundation; the Italian
Istituto Nazionale di Fisica Nucleare; the Ministry of Education, Culture,
Sports, Science and Technology of Japan; the Natural Sciences and Engineering
Research Council of Canada; the National Science Council of the Republic of
China; the Swiss National Science Foundation; the A.P. Sloan Foundation;
the Bundesministerium f\"ur Bildung und Forschung, Germany; the Korean
Science and Engineering Foundation and the Korean Research Foundation;
the Particle Physics and Astronomy Research Council and the Royal Society,
UK; the Institut National de Physique Nucleaire et Physique des
Particules/CNRS; the Russian Foundation for Basic Research; the
Comisi\'on Interministerial de Ciencia y Tecnolog\'{\i}a, Spain;
the European Community's Human Potential Programme under contract
HPRN-CT-2002-00292; and the Academy of Finland.


\begin{thebibliography}{99}
\bibitem{charge conjugate}
Charge conjugation is assumed throughout this report and we 
omit the symbols to identify particles and anti-particles.
For example, $W^{\prime+}\to e^+\nu_{e}$ and its charge
conjugate are denoted as $W^\prime\to e\nu_{e}$.
\bibitem{met}
Missing transverse energy, $\met$, is defined as the magnitude of 
$-\sum_{i}E_T^i\hat{n}_i$, where $\hat{n}_i$ is a unit vector in the 
azimuthal plane that points from the beam line to the $i$th 
calorimeter tower.
\end{thebibliography}
\end{document}